%% file: main.tex
\documentclass[conference]{IEEEtran}
\usepackage{times}

\usepackage[numbers]{natbib}
\usepackage{multicol}
\usepackage[bookmarks=true]{hyperref}

\usepackage{graphicx}          
\makeatletter \let\cl@part\relax \makeatother 
\usepackage{amsmath} 
\usepackage{amssymb}
\usepackage{balance}
\usepackage{mathtools}
\usepackage{siunitx}
\usepackage[usenames,dvipsnames]{xcolor}
\usepackage{pgfplots}
\usepgfplotslibrary{fillbetween}
\usepackage{tabularx}
\usepackage{booktabs}
\pgfplotsset{compat = newest}
\usetikzlibrary{arrows,positioning,shapes,intersections,patterns,calc}

\input{mydefs.tex}

\begin{document}

\title{Keep soft robots soft - a data-driven based trade-off between feed-forward and feedback control}

\author{Thomas Beckers and Sandra Hirche\\Chair of Information-oriented Control (ITR),
Department of Electrical and Computer Engineering,\\Technical University
of Munich, 80333 Munich, Germany. \{t.beckers,hirche\}@tum.de}

\maketitle

\begin{abstract}
Tracking control for soft robots is challenging due to uncertainties in the system model and environment. Using high feedback gains to overcome this issue results in an increasing stiffness that clearly destroys the inherent safety property of soft robots. However, accurate models for feed-forward control are often difficult to obtain. In this article, we employ Gaussian Process regression to obtain a data-driven model that is used for the feed-forward compensation of unknown dynamics. The model fidelity is used to adapt the feed-forward and feedback part allowing low feedback gains in regions of high model confidence. 
\end{abstract}

\IEEEpeerreviewmaketitle

\vspace{-0.2cm}
\section{Introduction}
Soft robots represent one significant evolution of robotic systems, since they are designed to embody safe and natural behaviors~\cite{della2017controlling}. The control of the robots is often challenging due to complex physical structures based on the soft materials. A common approach is to derive a simple, dynamic model from first order physics and increase the feedback gains to compensate the uncertainties until a desired tracking performance is achieved~\cite{spong2006robot}. However, high gain control is undesirable since this results in an increased stiffness that destroys the inherent safety properties~\cite{della2017controlling}; deriving a more accurate model of the system is often difficult if not impossible~\cite{amiri2016control}.\\
To overcome this issue, data-driven techniques deliver promising results in modeling and control of soft robot dynamics~\cite{reinhart2017hybrid,gillespie2018learning,bruder2019modeling}. The accurate model based on data-driven techniques allows a more precise feed-forward control such that high feedback gains are needless. However, it is often difficult to decide if the quality of the model is good enough to rely on feed-forward control only or if feedback control is required to ensure the tracking performance.\\
Gaussian Process regression (GPR) is a supervised learning technique which provides not only a mean function but also a predicted variance, and therefore a measure of the model fidelity based on the distance to the training data. It requires only a minimum of prior knowledge, generalizes well even for small training data sets and has a precise trade-off between fitting the data and smoothing~\cite{rasmussen2006gaussian}. \\ 
In this article, we present a GPR based control law for soft robots with automatic trade-off between feed-forward and feedback control. For this purpose, a GP learns the unknown system dynamics from training data. The proposed control law uses the mean of the GPR to compensate the unknown dynamics in a feed-forward manner and the model fidelity to adapt the influence of the feedback control part. As result, the the robot is softer in regions with training data.
\section{Methodology}
We focus on the class of pneumatic- or tendon-actuated, worm-like robots. For the modeling, the robot is virtually separated into $n_s\gg 1$ rigid sections with constant curvature, see e.g. \cite{olguin2018modelling}. Each section consists of $n_a$ actuators, e.g. the number of muscles. \Citet{falkenhahn2015dynamic,della2018dynamic} develop for such class of robots an approximated dynamical model given by
\begin{align}
\bm{\tau}(\q,\p)=M(\q)\ddq+C(\dq,\q)\dq+N(\dq,\q),\,\y=\g(\q),\label{for:dyn_model}
\end{align} 
where $\q\in\R^{n},n=n_s n_a$ is the generalized vector of positions and $\y\in\R^m$ the output of the system defined by the sensors used, e.g. a vision based system~\cite{fang2019vision}. In addition, there are the positive definite mass matrix $M(\q)\in\R^{n\times n}$, the Coriolis matrix $C(\dq,\q)\in\R^{n\times n}$ and the force vector $N(\dq,\q)\in\R^{n}$. The passive forces of gravity and the force resulting from the nominal position of the robot are included in $N$. The actuator force vector $\bm{\tau}(\q,\p)\in\R^n$ depends on the applied force $\p$ along the generalized coordinates and the muscle lengths $\q$. 
\subsection{Learning}
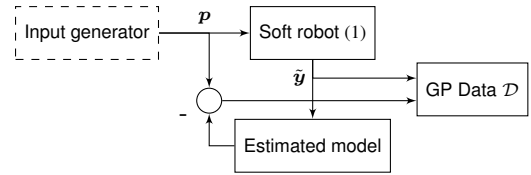
\begin{figure}[h]
	\begin{center}
	 \vspace{-0.2cm}	
	 	\input{figure/training_bsb.tex}
	 	\vspace{-0.2cm}
		\caption{Generation of the training data set for the GP.\label{fig:TD}}
	\end{center}
	\vspace{-0.4cm}
\end{figure}

To achieve a controller with a good feed-forward compensation, the system~\cref{for:dyn_model} must be identified. Since partial a priori knowledge of the system is often available, we use a gray-box model which combines an estimated and a data-driven model, see~\cite{reinhart2017hybrid}. A Gaussian Process learns the difference between the actual and the estimated output given by $\hat\p=\bm{h}(\tilde{\y})$, where $\tilde{\y}=[\ddot{\y}^\top,\dot{\y}^\top,\y^\top]^\top,\hat\p\in\R^n$ is the estimated force for the output $\tilde{\y}$. For this purpose, a set~$\mathcal D=\{Y,P\}$ consisting of~$n_d$ data points~$Y=[\tilde{\y}^{\{1\}},\ldots,\tilde{\y}^{\{n_d\}}]\in\R^{3m\times n_d}$ and~$Y=[(\hat\p-\p)^{\{1\}},\ldots,(\hat\p-\p)^{\{n_d\}}]\tran\in\R^{n\times n_d}$ is recorded, as visualized in~\cref{fig:TD}. The function $\bm{h}\colon\times\R^{3m}\to\R^n$ of the estimated model can be determined by system identification or first order principles.
If no a priori knowledge is available, the function $\bm{h}$ is set to zero. A force generator produce the desired profiles to collect at least a finite number of training points. After the data collecting step, the generated data is used to train a GP model.
\subsection{Control}
The control of the soft robot consists of a feed-forward and a feedback part. For the desired output $\tilde{\y}_d=[\ddot{\y}_d^\top,\dot{\y}_d^\top,\y_d^\top]^\top$, the predicted mean of the Gaussian process
\begin{align*}
	\mean(p_i\vert \tilde{\y}_d,\mathcal D)&=\bm k(\y_d,Y)\tran (K+I \sigma^2_{i})^{-1}P_{:,i}
\end{align*}
combined with the estimated model are used to estimate the necessary force $\p$ that is used as feed-forward control. The prediction of each component of~$\p$ for a $\tilde{\y}_d$ is derived from a Gaussian joint distribution with the covariance function~$k\colon\R^n\times\R^n\to\R$ as a measure of the correlation of two points~$(\tilde{\y},\tilde{\y}^\prime)$, see~\cite{rasmussen2006gaussian}. Alternatively, a vector-valued GP regression could also consider the spatial correlations among the input values~\cite{liu2018remarks}. The matrix function~$K\colon\R^{n\times n_d}\times \R^{n\times n_d}\to\R^{n_d\times n_d}$ is called the covariance matrix $K_{j,l}= k(Y_{:, l},Y_{:, j})$ where each element of the matrix represents the covariance between two elements of the training data~$Y$. The vector-valued covariance function~$\bm k\colon\R^n\times \R^{n\times n_d}\to\R^{n_d}$ calculates the covariance between~$\tilde{\y}_d$ and the training data~$Y$, i.e. $k_{j} = k(\tilde{\y}_d,Y_{:, j})$ for all~$j$. These functions depend on a set of hyperparameters that parametrize the covariance function. The variance $\sigma_{i}$ of the output noise can be determined based on the output data $P$. A comparison of the characteristics of the different covariance functions and tuning algorithms for the hyperparameters can be found in~\cite{bishop2006pattern}.\\
A feedback controller $\p=\bm{u}(\y)$ with $\bm{u}\colon \R^m\to\R^n$, e.g. a PID-controller, should eliminate the remaining control error due to model uncertainties. 
The feed-forward control allows to keep the robot soft, see~\cite{della2017controlling}. Therefore, the feedback control part should only be used if the feed-forward generated force is not accurate enough. To monitor the quality of the estimated force, we use the predicted variance 
\begin{align*}
	\var(p_i\vert \tilde{\y}_d,\mathcal D)\!=\!k(\tilde{\y}_d,\tilde{\y}_d)\!-\!\bm k(\tilde{\y}_d,Y)\tran (K+I \sigma^2_{i})^{-1} \bm k(\tilde{\y}_d,Y)
\end{align*}
of the Gaussian Process. Based on this variance, the weight of the feedback and feed-forward part is adapted by a function $\alpha\colon\R^n\to [0,1]$. Thus, the control law is given by
\begin{align*}
\begin{split}
\p&=\Big(1-\alpha\big(\Var(\hat\p\Vert \tilde{\y}_d,\mathcal D)\big)\Big)\big(\underbrace{\Mean(\hat\p\Vert \tilde{\y}_d,\mathcal D)+\bm{h}(\tilde{\y}_d)}_{\text{Feed-forward}}\big)\\
&+\alpha\big(\Var(\hat\p\Vert \tilde{\y}_d,\mathcal D)\big)\underbrace{\bm{u}(\y_d-\y)}_{\text{Feedback}},
\end{split}
\end{align*}
where $\bm \mean(\p\vert \tilde{\y}_d,\mathcal D)=[\mean(p_1\vert \tilde{\y}_d,\mathcal D),\ldots,\mean(p_n\vert \tilde{\y}_d,\mathcal D)]\tran$ and $\Var(\p\vert \tilde{\y}_d,\mathcal D)=\diag(\var(p_1\vert \tilde{\y}_d,\mathcal D),\ldots,\var(p_n\vert \tilde{\y}_d,\mathcal D))$. A possible function for $\alpha$ is, e.g. the sigmoid function
\begin{align}
\alpha(\Var)=\sig\big(c_1\Var(\hat\p\Vert \tilde{\y}_d,\mathcal D)+c_2\big),\quad c_1,c_2\in\R.\label{for:tan}
\end{align}
Thus, in regions with high model fidelity, the feedback is reduced whereas an uncertain force prediction increase the weight of the feedback to keep the control error low. As result, the robot remains softer while keeping the tracking error low. A similar control law for Euler-Langrange systems with guaranteed stability is presented in~\cite{beckersauto2018}.
\section{Simulation}
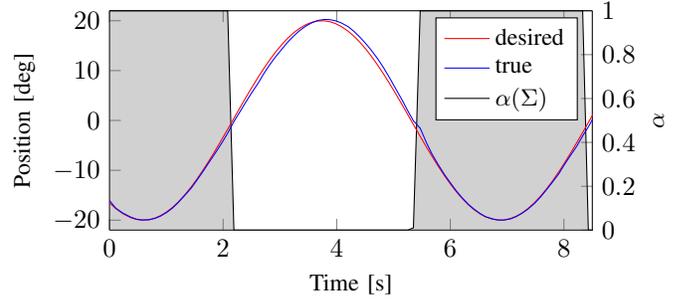
\begin{figure}[h]
	\begin{center}
	 \vspace{-0.3cm}	
	 	\input{include/plot_error.tex}
	 	\vspace{-0.8cm}
		\caption{Small tracking error inside and outside the training data set. The gray area indicates a high $\alpha$-value and thus, feedback control.\label{fig:tracking}}
	\end{center}
	\vspace{-0.6cm}
\end{figure}

\begin{figure}[h]
	\begin{center}
	 	\includegraphics[width=0.7\columnwidth]{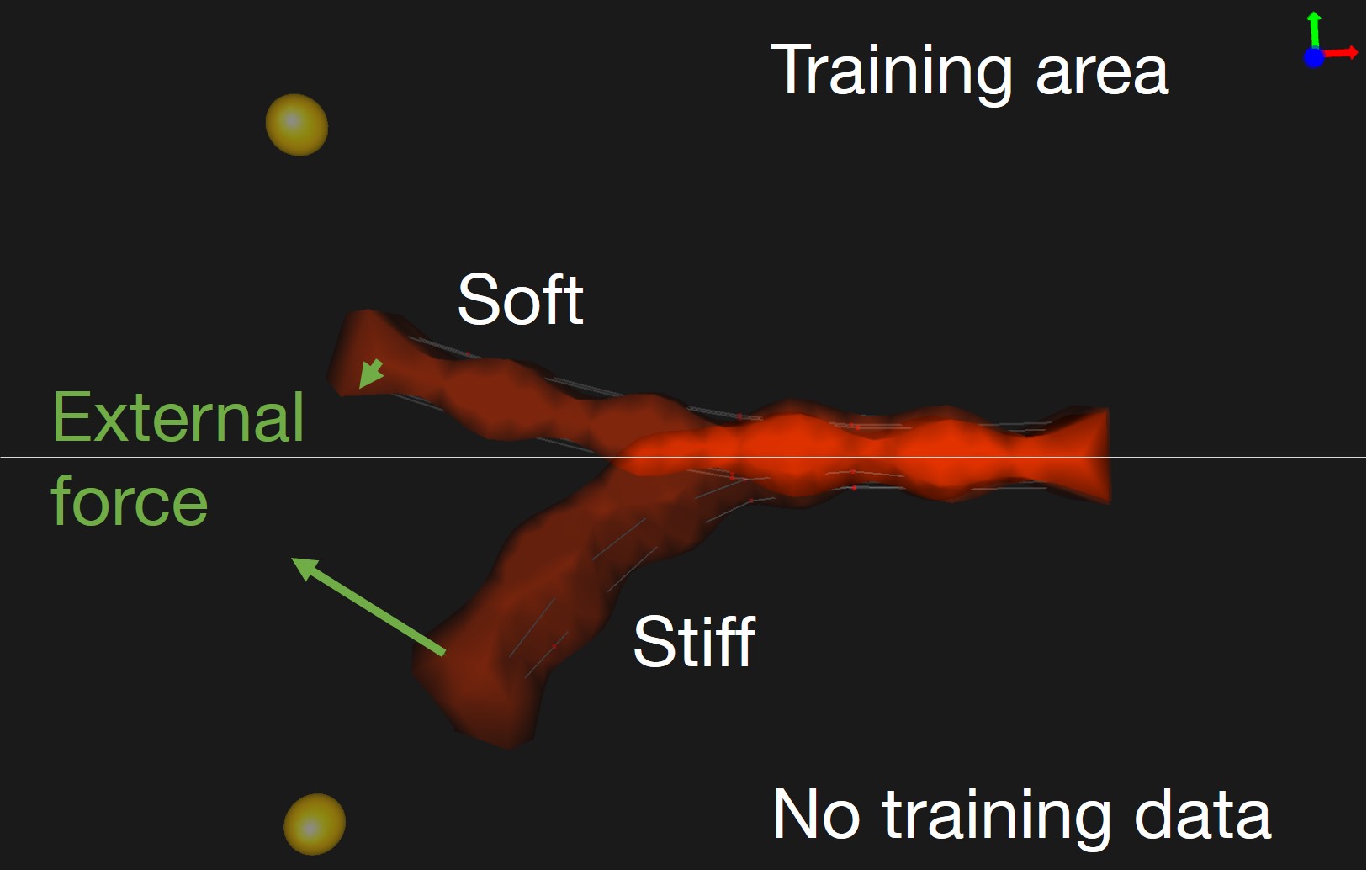}
	 	\vspace{-0.2cm}
		\caption{The robot remains soft in regions of high model confidence because of GP based feed-forward control. Video: \url{https://youtu.be/vArJZukAzDA}\label{fig:robot}}
	\end{center}
	\vspace{-0.2cm}
\end{figure}

A simulation with the proposed control law is implemented in Matlab/Simulink and the soft robot framework SOFA~\cite{Faure2012}. A worm-like robot is modeled with one artificial muscles on each side. The output $y$ of the robot is the angle of the tip with respect to the base. 250 data points consisting of the tip position and the force are collected in the upper half plane of the work space are used to train a GP with squared exponential kernel. The hyperparameters are optimized based on the likelihood function. The estimated model is set to zero and the feedback part is realized with a discrete PI-controller.\\ 
Depending on the predicted variance of the GP, a sigmoid function as~\cref{for:tan} is used to weight between feed-forward and feedback control. Although the undisturbed system's tracking error is small inside and outside the training area~\cref{fig:tracking}, there is a significant difference regarding the stiffness~\cref{fig:robot}. 
Inside the training area, the robot is soft which is shown by a large deflection due to a small force (green arrow). In contrast, outside the training area, the stiffness is increasing due to the feedback control to keep the tracking error low.
\section{Conclusion} 
\label{sec:conclusion}
We present a Gaussian Process based control law that keeps the robot soft without losing the tracking performance. For this purpose, an automatic weighting between feed-forward and feedback control based on the predicted variance of the GP is proposed. If the the robot operates inside an area of the workspace where training data was collected, the predicted mean of a GP is used as feed-forward control. Outside this area, the feedback part is getting more dominant that keeps the tracking error low but also increase the stiffness.

\section*{Acknowledgments}
The research leading to these results has received funding from the ERC Starting Grant ``Control based on Human Models (con-humo)'' agreement n\textsuperscript{o}337654.

\bibliographystyle{plainnat}
\bibliography{references}

\end{document}

%% file: mydefs.tex
\newcommand\tran{\mkern-2mu\raise1.25ex\hbox{$\scriptscriptstyle\top\hspace{0.5mm}$}\mkern-3.5mu}
\newcommand{\R}{\mathbb{R}}

\newcommand{\bm}[1]{{\boldsymbol{#1}}}

\DeclareMathOperator{\diag}{diag}
\DeclareMathOperator{\sig}{sig}
\DeclareMathOperator{\var}{var}
\DeclareMathOperator{\mean}{\mu}
\DeclareMathOperator{\Var}{\Sigma}
\DeclareMathOperator{\Mean}{\bm\mu}

\newcommand{\g}{\bm g}

\newcommand{\ddq}{\ddot{\bm q}}
\newcommand{\dq}{\dot{\bm q}}
\newcommand{\q}{\bm q}
\newcommand{\p}{\bm p}

\newcommand{\y}{\bm y}
\usepackage[noabbrev]{cleveref} 
\crefname{propy}{Property}{Properties}
\crefname{rem}{Remark}{Remarks}
\crefname{assum}{Assumption}{Assumptions}
\crefname{prop}{Proposition}{Propositions}
\crefname{cor}{Corollary}{Corollaries}
\crefname{lem}{Lemma}{Lemmas}
\crefname{thm}{Theorem}{Theorems}
\crefname{defn}{Definition}{Definitions}
\crefname{figure}{Fig.}{Fig.}
\Crefname{figure}{Figure}{Figures}
\crefname{equation}{}{}

%% file: figure/training_bsb.tex
\begin{tikzpicture}[auto, node distance=3cm,>=latex']
\tikzstyle{block} = [draw, fill=white, rectangle, minimum height=2em, minimum width=1em]
\tikzstyle{sum} = [draw, fill=white, circle, node distance=1cm]
\tikzstyle{input} = [coordinate]
\tikzstyle{output} = [coordinate]
\tikzstyle{t_output} = []
\tikzstyle{pinstyle} = [pin edge={to-,thin,black}]

    \node [block, dashed, name=input] {$\scriptstyle \textsf{Input generator}$};
    \node [block, right of=input,node distance=3cm] (robot) {$\scriptstyle \textsf{Soft robot } \cref{for:dyn_model}$};
    \node [block, below of=robot,node distance=1.5cm] (hatrobot) {$\scriptstyle \textsf{Estimated model}$};

	\draw [->] (input) -- node[name=t,above] {$\scriptstyle \p$} (robot);
    \draw [->] (robot) -- node[name=q,label={[xshift=-0.3cm, yshift=-0.2cm]$\scriptstyle \tilde{\y}$}] {} (hatrobot);
    
    \node [sum, left of=q,yshift=-0.15cm,node distance=1cm,label={[label distance=0cm]200:-},node distance=1.5cm] (sum) {};
    
    \node [block, right of=q,node distance=2cm] (x_output) {$\scriptstyle \textsf{GP Data }\mathcal{D}$};
    
    \draw [->] (input) -| (sum);
    \draw [->] (hatrobot) -| (sum);
    \draw [->] (sum) -- ([yshift=-0.15cm]x_output.west);
    \draw [->] (robot) |- ([yshift=0.15cm]x_output.west);

\end{tikzpicture}

%% file: include/plot_error.tex
\begin{tikzpicture}
\pgfplotsset{
  set layers,
  mark layer=axis tick labels
}
\pgfplotsset{select coords between index/.style 2 args={
    x filter/.code={
        \ifnum\coordindex<#1\def\pgfmathresult{}\fi
        \ifnum\coordindex>#2\def\pgfmathresult{}\fi
    }
}}
\begin{axis}[
axis y line*=right,
 ylabel={\small $\alpha$},
    height=4.5cm,
    width=8cm,
  xmin=0, xmax=8.5, ymin=0, ymax=1,
  axis x line=none]
\addplot+[name path=varp1, color=black, no marks] table [x index=0,y index=3]{data/data.dat};
\addplot+[name path=varp2, color=black!60!white,opacity=0.3, no marks] table [x index=0,y index=4]{data/data.dat};
\addplot[color=black!60!white,opacity=0.3] fill between[of=varp1 and varp2];
\end{axis}
\begin{axis}[
axis y line*=left,
  xlabel={\small Time [s]},
  ylabel={\small Position [deg]},
    height=4.5cm,
    width=8cm,
  xmin=0, xmax=8.5, ymin=-22, ymax=22,
  legend pos=north east,
  legend cell align={left}]
\addplot+[color=red,no marks] table [x index=0,y index=1]{data/data.dat};
\addplot+[color=blue,no marks] table [x index=0,y index=2]{data/data.dat};
\addplot+[color=black, no marks] coordinates {(100,100)};
\legend{\small desired, \small true, \small $\alpha(\Sigma)$}
\end{axis}
\end{tikzpicture} 